\documentclass[a4paper,11pt]{article}
\usepackage{pos}

\usepackage{wrapfig}
\usepackage{xspace}
\usepackage{multicol}
\usepackage{enumitem}

\title{AugerPrime: Status and first results}

\author*[a]{David Schmidt}
\onbehalf{for the Pierre Auger Collaboration${}^b$} 

\affiliation[a]{Institute for Astroparticle Physics, Karlsruhe Institute of Technology (KIT)\\
  Kaiserstraße 12, Karlsruhe, Germany}
\affiliation[b]{Observatorio Pierre Auger, Av.\ San Mart{\'\i}n Norte 304, 5613 Malarg\"ue, Argentina\\
Full author list: {\rm\url{https://www.auger.org/archive/authors_icrc_2025.html}}}

\emailAdd{spokespersons@auger.org}

\abstract{With the knowledge and statistical precision derived from two decades of measurement, the Pierre Auger Observatory has significantly deepened our understanding of ultra-high-energy cosmic rays while unearthing an increasingly complex astrophysical landscape and exposing tensions with hadronic interaction models.
The field now demands the mass of individual cosmic-ray primaries as an observable with an exposure that only the 3000-square-kilometer surface array of the Observatory can provide.
Access to the primary mass hinges on the disentanglement of the electromagnetic and muonic components of extensive air showers.
To achieve this, scintillator and radio detectors have been installed atop each existing water-Cherenkov detector of the surface array, whose dynamic range has also been enhanced through the installation of small-area PMTs.
Additionally, the timing and signal resolution of all detector stations have been improved through upgraded station electronics, and underground muon counters have been installed in a region of the array with denser spacing.
As the commissioning of the final components of AugerPrime reaches its conclusion and the enhanced array comes fully online, we present the realization of its design, its performance, and the first results from this now multi-hybrid observatory.}

\ConferenceLogo{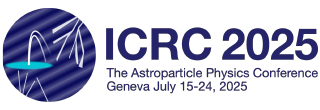}

\FullConference{39th International Cosmic Ray Conference (ICRC2025)\\
 15–24 July 2025\\
Geneva, Switzerland\\}

\begin{document}
\maketitle

\section{Introduction}

Using the reconstructed energies and arrival directions of cosmic rays, the 3000\,km$^2$ surface detector array of the Pierre Auger Observatory -- with its exposure now exceeding 100\,000\,km$^2$\,sr\,yr -- has enabled the establishment of anisotropy in the arrival directions of ultra-high-energy cosmic rays (UHECRs) with a significance exceeding 5$\sigma$ for the first time ever\,\cite{PierreAuger:2017pzq,PierreAuger:2024fgl}.
Notably, this result was obtained without any information about the masses of the primary cosmic rays.
In parallel, measurements of the depth of shower maximum $X_\mathrm{max}$ by the fluorescence detector of the Pierre Auger Observatory have transformed understanding of UHECR composition \cite{PierreAuger:2014sui,PierreAuger:2014gko}.
Contrary to earlier assumptions that the flux at the highest energies was dominated by protons, current evidence indicates a trend toward increasingly heavier nuclei with rising energy.
To probe more deeply into the astrophysical origins of the established and increasingly significant anisotropies in arrival directions, it is clear that studies with enhanced mass sensitivity are essential.
It is also clear that the exposure required for such studies at these energies exceeds that of the fluorescence detector of the Observatory and must come from its surface detector.

The application of machine learning techniques shows promise to be a revolution in the reconstruction of the properties of primary cosmic rays from surface detector measurements.
The efficacy of these techniques for reconstructing $X_\mathrm{max}$ in individual surface detector events has been demonstrated \cite{PierreAuger:2024nzw} revealing previously unobserved breaks in the evolution of $X_\mathrm{max}$ with energy \cite{PierreAuger:2024flk} -- features that could not be significantly resolved using the fluorescence detectors alone due to their limited exposure.
Despite their potential, the ability to confidently employ machine learning algorithms is currently limited to predicting a small subset of observables, however, as the simulations of air showers upon which deep neural networks (DNNs) are trained have known differences with real air showers.
These differences result in biases visible when comparing with direct measurements of the predicted quantities.
In the case of $X_\mathrm{max}$, the observed bias in DNN predictions is approximately 30\,g/cm$^2$, which has been dealt with by calibrating with fluorescence detector measurements.
To fully harness the power of neural networks in combination with the extensive exposure of the surface detector of Auger, it is essential that the observables predicted by these networks are either calibrated using a reliable set of direct measurements or that the hadronic interaction models used to generate the simulated air showers upon which networks are trained are improved to more closely reproduce observational data.
Each of these points requires measurement of the individual components of extensive air showers.
Understanding the mismatch between simulated air showers and measurements also exists as a scientific objective of high interest in its own right.

The AugerPrime upgrade to the Pierre Auger Observatory will deliver a data set of UHECR measurements with mass as an observable for primary cosmic rays incident on the array with nearly all inclinations.
Known as Phase II, the 3000\,km$^2$ surface detector of the observatory will be operated in the AugerPrime configuration for a minimum of 10 years (i.e., until at least 2035).
In addition to delivering a prime data set for mass enhanced studies of anisotropy in the arrival directions of UHECR, Phase II will include detailed measurements of the different components of extensive air showers on a shower-by-shower basis up to the highest energies, which will serve to inform hadronic interaction models and define the mass scale of the observatory's measurements with increased confidence.
It will also allow for the re-analysis and calibration of observables predicted with advanced methods, such as neutral networks, for measurements with the enormous exposure of Phase I of the observatory -- where only the water-Cherenkov detectors were deployed for the surface detector array -- as well as directly improve the precision and accuracy of the predictions of such algorithms by meaningfully contributing to the alignment of the simulations they are trained upon with the properties of real air showers.

\section{Basic design}

\begin{figure}
    \centering
    \includegraphics[width=0.44\linewidth]{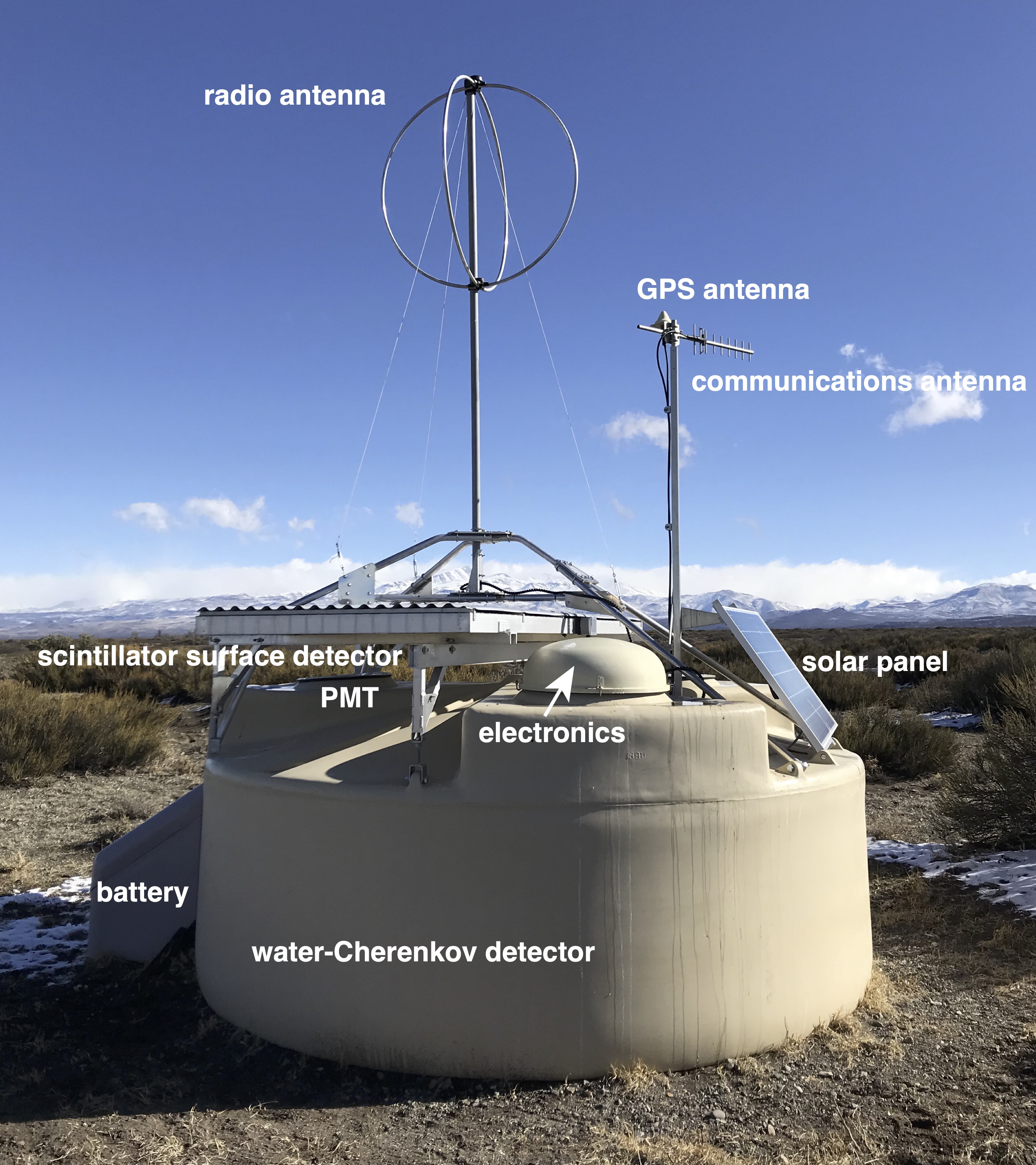}\hfill
    \includegraphics[width=0.545\linewidth]{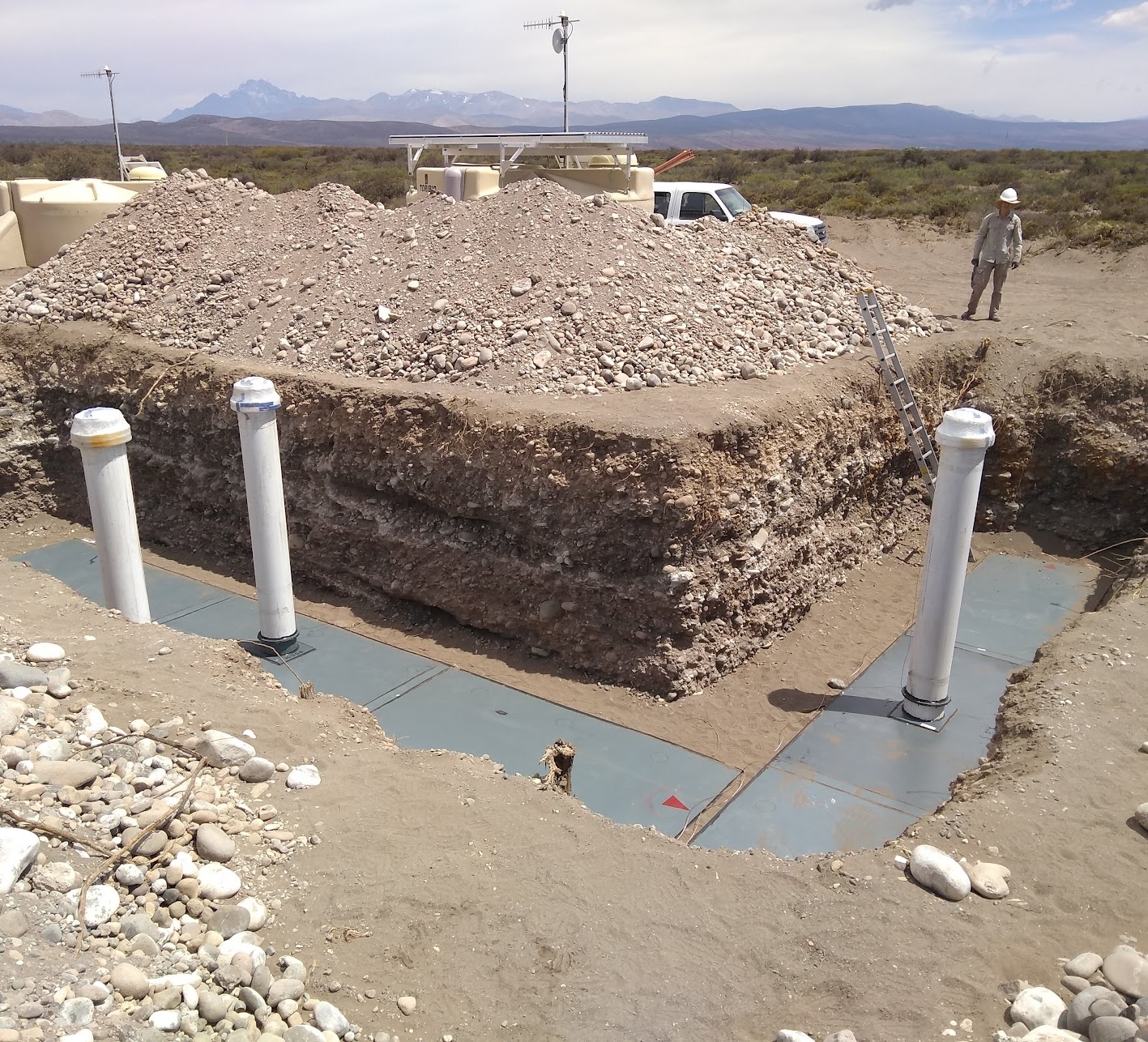}
    \caption{\textit{Left}: A fully deployed AugerPrime (Phase II) surface detector station. \textit{Right}: A counter of the Underground Muon Detector during deployment.}
    \label{fig:augerprime_station}
\end{figure}

The objective of the AugerPrime upgrade is to enhance the mass sensitivity of the surface detector array by adding detectors with complementary responses to the different components of extensive air showers, while maximizing the sky coverage over which this sensitivity is achieved.
This is achieved by installing planar scintillation-based detectors -- referred to as Scintillator Surface Detectors (SSDs) -- along with radio antennas -- collectively referred to as the Radio Detector (RD) -- on top of the existing water-Cherenkov detectors (WCDs) that form the 1500\,m-spaced isometric triangular grid of the 3000 km$^2$ surface detector array.
An image of an upgraded surface detector station is shown in Fig.\,\ref{fig:augerprime_station}-left.

The addition of the SSD provides mass sensitivity in that its response to the electromagnetic and muonic shower components differs from that of the WCD.
This allows for a deconvolution to obtain the magnitudes of the contribution of each component to the total detector signals, or alternatively, the inclusion of the time-dependent signals of each detector into the global likelihoods of more complex reconstruction algorithms.
For more inclined air showers, i.e., those with a zenith angle $\theta \gtrsim 60^\circ$, the scintillator measurements are less effective due to the extensive attenuation of the electromagnetic shower component and the detectors' decreased projected area in the shower plane.
At these higher inclinations, however, the radio footprint of air showers is sufficiently large at the ground such that the energy of showers can be effectively estimated from its sampling.
With the energy of the shower in hand, the WCDs provide the mass-sensitivity for these inclined showers.
In this way, mass sensitivity is achieved for effectively the full sky observed with Auger -- by the addition of scintillators for less inclined showers and by the addition of radio antennas for more inclined showers.

A 20\,km$^2$ sector of the surface array with 750\,m spacing, including a 2\,km$^2$ sub-sector with 433\,m spacing, is being instrumented with additional scintillation-based detectors buried at each position -- collectively referred to as the Underground Muon Detector (UMD).
These shielded scintillators allow for the direct measurement of the muon component of extensive air showers for inclinations up to approximately 60$^\circ$ in zenith angle.
In addition to delivering a data set with high-precision muon measurements for air showers from cosmic rays with energies between the second knee and the ankle of the cosmic ray energy spectrum, the direct muon measurements will also serve to calibrate algorithms used to estimate muon content with measurements of the upgraded stations of the 3000\,km$^2$ array.

\section{Realization}

\subsection{Scintillator Surface Detector}

An SSD module has an active detection area of 3.84\,m$^2$, which is comprised of 48 extruded polystyrene scintillator bars each measuring 1.6\,m in length and 5\,cm in width with a thickness of 1\,cm.
The bars are distributed between two planes with an aluminum tube housing a 1.5-inch diameter Hamamatsu R9420 PMT set between them.
Light produced in the bars is routed to the PMT through wavelength-shifting fibers passing through two holes in each scintillator bar.
The active components of the SSDs are housed inside an aluminum enclosure.
The design of the mechanical structure of the SSD ensures that it is light-tight, sufficiently rigid for transportation, and sufficiently robust to withstand more than 10 years of operation in the field.
Additionally, the modules feature a sunroof made of corrugated aluminum to mitigate temperature variations.
More details on the detector design can be found in \cite{Smida:2017zmh}.
The SSD modules were assembled and underwent extensive quality controls at six sites in different Auger institutions \cite{Pkala:2020cro}.
Two additional institutions were dedicated to testing and preparing the PMTs.

A single minimum ionizing particle (MIP) in the SSDs produces around 30 photoelectrons, and the dynamic range of the SSD PMT and electronics allows measurement up to and exceeding 20\,000 (MIPs) with no more than 5\% deviation from a linear response.
The SSDs are calibrated using the continuous background of atmospheric particles by fitting the ``muon hump'' in the distribution of charge they produce and making use of the experimentally validated relationship of this hump position with the charge of a vertical MIP.
Details on this process and on the performance of the SSDs may be found in \cite{ICRC25:Matteo, ICRC25:Belen}.

\subsection{Radio Detector}

The dual-polarized short aperiodic loaded loop antennas of the RD consist of two aluminum rings with diameters of 122\,cm.
The rings are oriented perpendicular to one another and are respectively aligned parallel and perpendicular to the Earth's magnetic field.
The antennas are fixed to a mast attached to an aluminum frame mounted directly on the WCDs with guy-wires providing additional support and mitigation of vibrations induced by strong winds.

The sensitivity of the antennas lies in the frequency range of 30 to 80\,MHz in which its response is virtually uniform with low dispersion.
The 12-bit, 250\,MHz electronics have an amplification of a total of 36 dB and include a band-pass filter in the 30 to 80 MHz range.
An FPGA coordinates data exchange with the station electronics discussed below.

Calibration of the radio antennas has its basis in a thorough understanding of the full signal chain through laboratory measurements and simulations.
The directional response of the full chain is validated through measurements with a radio calibration source mounted on a drone, and the radio emission from the Milky Way serves as an absolute calibration source which has thus far been used to validate the absolute gain of the signal chain.
Additional information on the calibration and performance of the RD can be found in \cite{ICRC25:Bjarni, ICRC25:Belen}.

\subsection{Electronics upgrade and extension of dynamic range}

\begin{figure}
    \centering
    \includegraphics[width=1\linewidth]{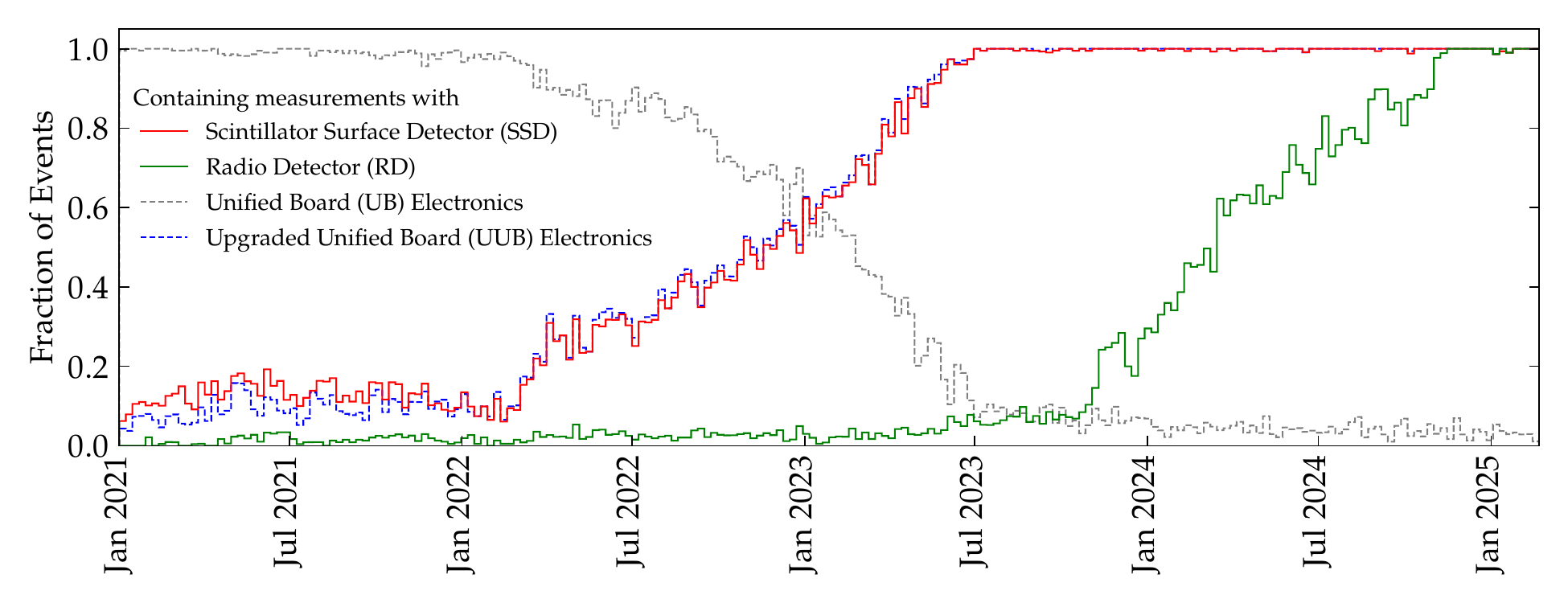}
    \caption{Fraction of events with energies greater than 10$^{18.5}$\,eV containing measurements with the different components of the AugerPrime upgrade during the transition period.}
    \label{fig:event_rate_transition}
\end{figure}

To match the dynamic range of the SSD, a 1-inch diameter Hamamatsu R8619 PMT -- referred to as the small PMT (SPMT) -- is installed in each WCD.
Whereas the WCD PMTs saturated at approximately 1000 vertical equivalent muons (VEMs) during Phase I of observatory operation, the dynamic range of the detectors is extended to approximately 20\,000\,VEM with the SPMT.

Individual muons are indistinguishable in the SPMT as they produce only about one photoelectron per muon.
Local, low energy showers are used for calibration with the rate of calibration events at approximately 200 per hour.
The differences between the signal spectra measured by the large and SPMTs in the WCDs is minimized for these showers to obtain a calibration for the SPMT.
Details on this procedure and performance of the SPMTs may be found in \cite{PierreAuger:2023clx, ICRC25:Belen,ICRC25:Martina}.

To accommodate the additional channels of the upgraded surface detectors and provide signal traces with enhanced temporal and signal resolution, the Unified Board (UB) electronics of Phase I were replaced with Upgraded Unified Board (UUB) electronics.
For the UUB, the anode channel inputs for each of the the large WCD PMTs are split and amplified such that the gain of the high-gain channel is 32 times the low-gain channel.
The anode channel of the SSD is also split with a ratio of 128 between the two gains.
The signals are filtered and digitized with a sampling frequency of 120\,MHz, which is three times that of the UB electronics.
The resolution on the magnitude of signals is also improved with the 12-bit UUB electronics, which improve upon the 10-bit UB.
The updated GPS receivers also boast an improved timing resolution of 5\,ns.
To accommodate the increased power consumption particularly due to the RD, new solar panels are also installed at each surface detector station.
An extensive description of the UUB electronics and their performance is given in \cite{PierreAuger:2023yab,ICRC25:Martina}.

\subsection{Underground Muon Detector}

\begin{figure}
    \centering
    \includegraphics[width=0.9\linewidth]{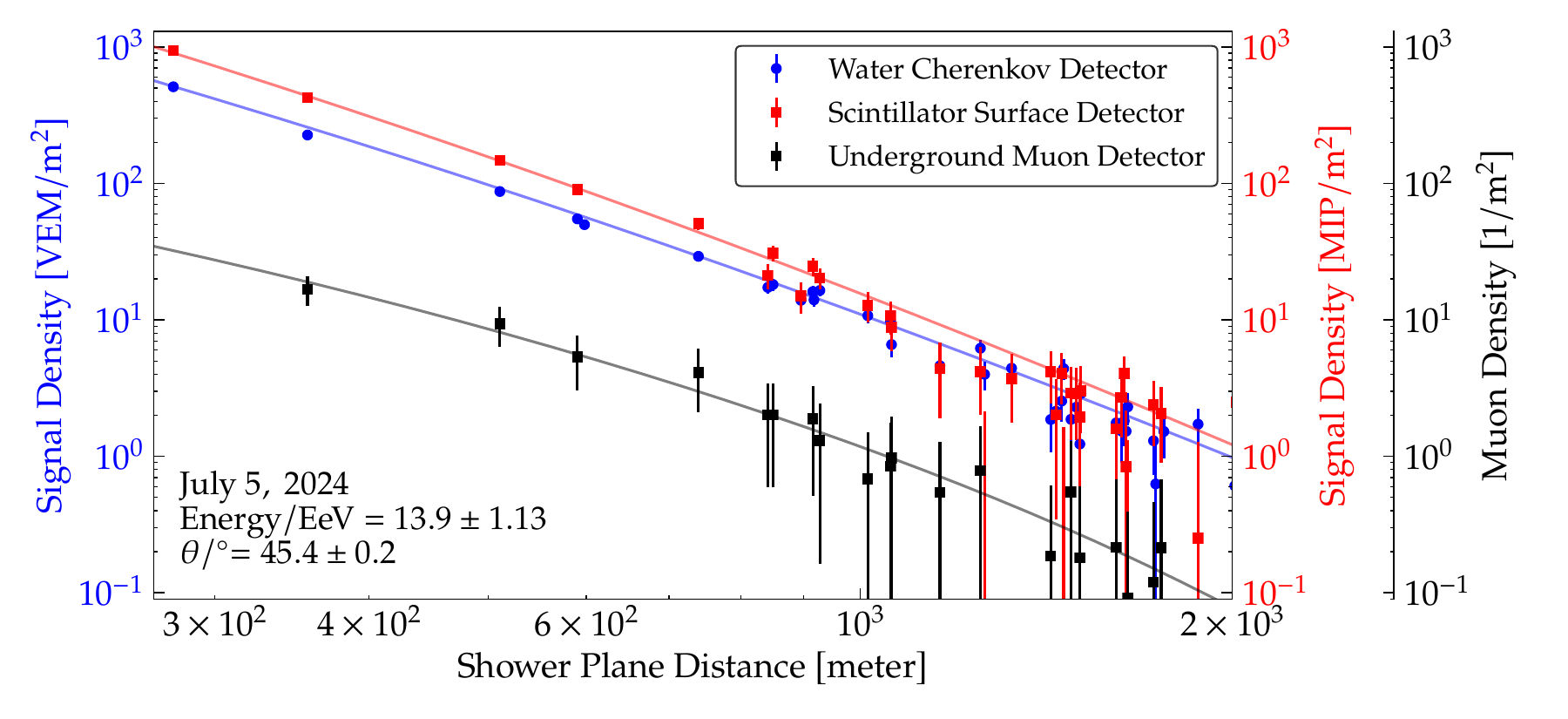}\hfill
    \caption{Lateral distributions of signals from the Water-Cherenkov Detectors, Scintillator Surface Detectors, and Underground Muon Detector for a sample Phase II event.}
    \label{fig:vertical_ldfs}
\end{figure}

A single position of the UMD consists of three scintillator modules each with an active area of 10.24\,m$^2$ resulting in a total active area of 30.72\,m$^2$.
A single module consists of 64 extruded polystyrene scintillator bars each measuring 4\,m in length and 4\,cm in width with a thickness of 1\,cm.
Wavelength-shifting fibers are embedded in each bar and route photons to a central array of 64 silicon photomultipliers.
The internal components of each module are enclosed in a water-tight, polyvinyl chloride casing and buried at a depth of 2.3\,m with a narrow access shaft for access to the electronics.
The overburden corresponds to 540\,g/cm$^2$ resulting in an energy threshold for muons at the ground to reach the buried detectors of approximately 1\,GeV.
The output of the SiPMs is processed with two read-out schemas.
An acquisition mode aimed at low muon densities applies a threshold to SiPM signals and generates 64 independent binary traces, i.e., one for each of the 64 scintillator bars.
The interpretation of these binary traces results in a muon count.
Details on the calibration of this acquisition mode using dark counts are given in \cite{PierreAuger:2020hrz}.
For higher muon densities, an additional acquisition mode sums the 64 traces and acts as a measurement of integrated charge \cite{ICRC25:Marina}.
More detailed information on the status and performance of the UMD is given in \cite{ICRC25:Joaquin,ICRC25:Belen}.

\begin{figure}
    \centering
    \includegraphics[width=0.49\linewidth]{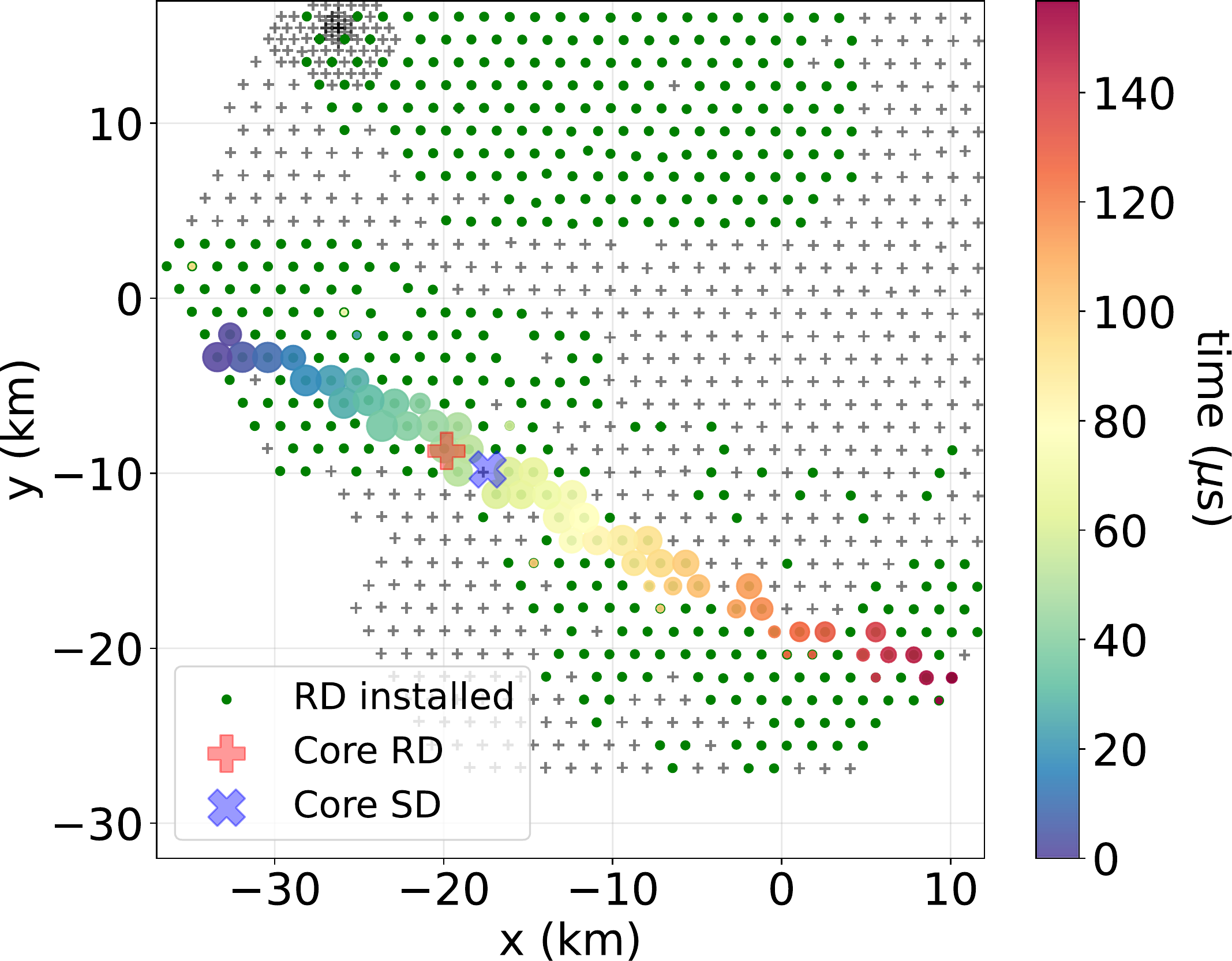}\hfill
    \includegraphics[width=0.48\linewidth]{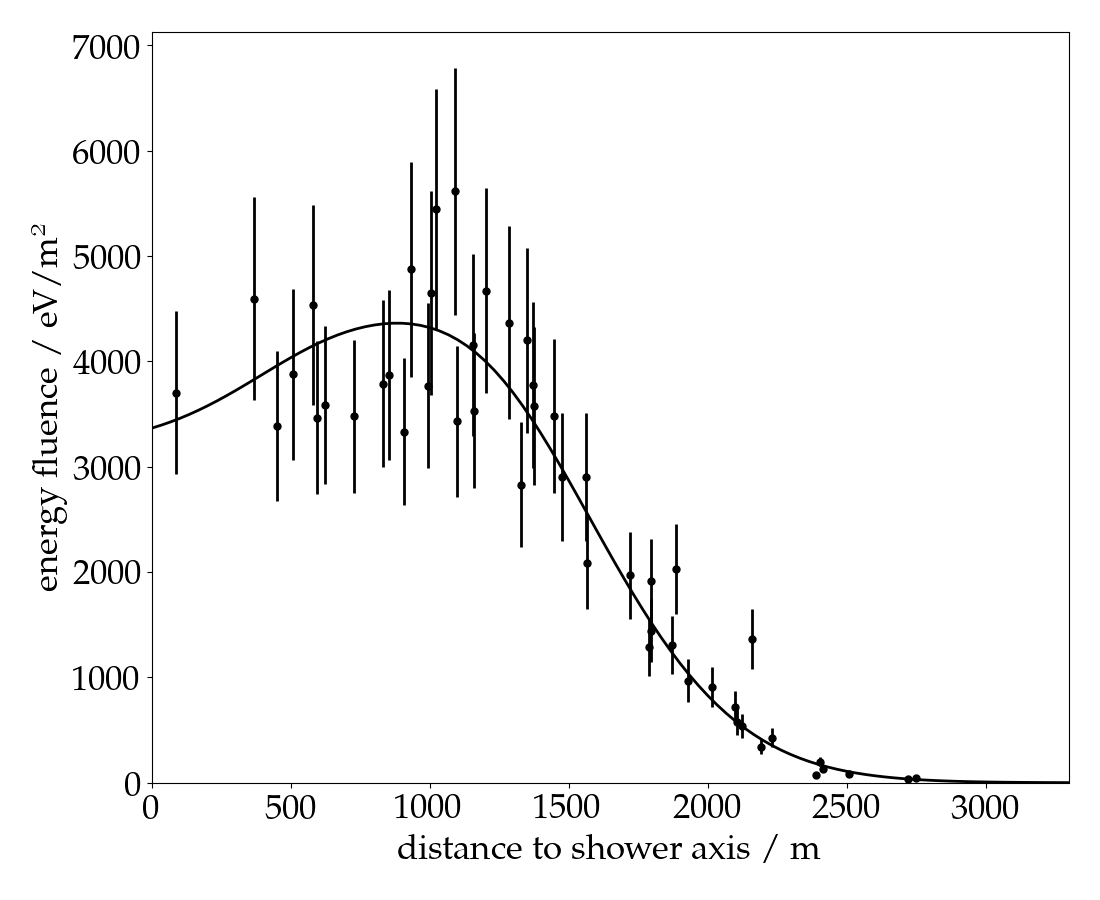}
    \caption{Topology (left) and lateral distribution (right) of energy fluence for a sample inclined event measured with the Radio Detector of Phase II.}
    \label{fig:rd_event}
\end{figure}

\section{Coming online}

The large scale deployment of the SSD modules began at the end of 2018 and was completed at the end of 2021.
The PMTs for the SSDs and the SPMTs were deployed thereafter together with the UUB electronics, the large scale deployment of which started with a pre-production batch in 2020 and was completed at the end of June 2023.
Each station in the surface detector array remained in acquisition until the moment its UB electronics were replaced with UUB electronics.
Upon the upgrade to its electronics, each station was then immediately put back into acquisition, now additionally providing SSD and SPMT measurements.
The transition of the surface detector from a WCD-only Phase I configuration to the upgraded Phase II configuration was therefore gradual, and the array operated in a mixed configuration during a period of slightly more than three years, as illustrated in Fig.\,\ref{fig:event_rate_transition}.
Currently, the UUBs operate in compatibility mode, in which the WCD signal traces are filtered and downsampled to emulate those of the original UB system for the purposes of triggering, allowing the application of Phase I trigger algorithms at the event level.
The native, full bandwidth traces of the UUBs are, however, used for the event reconstruction, and efforts towards the development of new, full bandwidth triggers aimed at the detection of neutral particles and making use of the new detectors of AugerPrime are on-going.
More details on the performance of the acquisition systems of the upgraded array are given in \cite{ICRC25:Paul}.
Exposure for the surface detector of Phase II is already approaching approximately 10\% of Phase I at the time of this proceeding.
An exemplary event including WCD, SSD, and UMD measurements is shown in Fig.\,\ref{fig:vertical_ldfs}.

RD deployment began in August of 2023 with the procurement of some components delayed through complications relating to the COVID-19 pandemic.
RD deployment was completed at the end of 2024.
Each RD antenna was put into acquisition upon its deployment.
An exemplary event measured during the RD deployment is shown in Fig.\,\ref{fig:rd_event}.

UMD mass production and deployment began in 2019.
At the time of this proceeding, 48 of 61 positions for the 20\,km$^2$ sector of the array with 750\,m spacing between stations have been deployed with deployment expected to be completed by the end of 2025.
The even denser sector of the array with 433\,m spacing between detectors has already been completed.

\section{Outlook}
As the number of Phase II events for multi-hybrid analysis steadily increases, the Auger collaboration is completing commissioning of the pipelines for physics analysis.
These efforts relate to calibration and monitoring of the upgrade detectors as well as interpreting the initial Phase II measurements, as discussed in  \cite{ICRC25:Marina,ICRC25:Belen,ICRC25:Matteo,ICRC25:Bjarni} in these proceedings.
Refinement of the reconstruction algorithms to be applied to Phase II measurements also continues, as presented in \cite{ICRC25:Simon,ICRC25:Bjarni,ICRC25:Darko,ICRC25:Harm}.

With AugerPrime now fully operational and the final steps of commissioning the physics data sets underway, a new era of analysis is beginning.
This includes mass-sensitive studies of anisotropies in arrival directions, improved precision in testing and constraining hadronic interaction models, and a deeper understanding of the mass composition of UHECRs -- extending to energies beyond the observed suppression in the spectrum.
In parallel, either measurements or significantly improved limits on neutral particles are anticipated, along with enhanced investigations into physics beyond the Standard Model.
In addition to the rich Phase II data set, insights gained from AugerPrime will also enable a re-analysis of Phase I data, leveraging its extensive exposure with greater confidence in the mass scale and a more refined understanding of the different components of extensive air showers.
New areas of study enabled by the upgraded detectors are also already beginning to emerge.
For example, ongoing efforts to quantify the lateral distribution and energy spectrum of neutrons in extensive air showers -- made possible for the first time by the addition of scintillators to the surface detector array -- may offer a novel window into hadronic interactions, as presented in \cite{ICRC25:Tobias} in these proceedings.

\newpage

\section*{The Pierre Auger Collaboration}

{\footnotesize\setlength{\baselineskip}{10pt}
\noindent
\begin{wrapfigure}[11]{l}{0.12\linewidth}
\vspace{-4pt}
\includegraphics[width=0.98\linewidth]{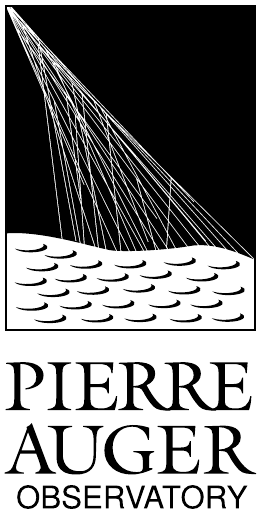}
\end{wrapfigure}
\begin{sloppypar}\noindent
A.~Abdul Halim$^{13}$,
P.~Abreu$^{70}$,
M.~Aglietta$^{53,51}$,
I.~Allekotte$^{1}$,
K.~Almeida Cheminant$^{78,77}$,
A.~Almela$^{7,12}$,
R.~Aloisio$^{44,45}$,
J.~Alvarez-Mu\~niz$^{76}$,
A.~Ambrosone$^{44}$,
J.~Ammerman Yebra$^{76}$
% ...

\end{sloppypar}
\begin{center}
\end{center}

\vspace{1ex}
\begin{description}[labelsep=0.2em,align=right,labelwidth=0.7em,labelindent=0em,leftmargin=2em,noitemsep,before={\renewcommand\makelabel[1]{##1 }}]
\item[$^{1}$] Centro At\'omico Bariloche and Instituto Balseiro (CNEA-UNCuyo-CONICET), San Carlos de Bariloche, Argentina
\item[$^{2}$] Departamento de F\'\i{}sica and Departamento de Ciencias de la Atm\'osfera y los Oc\'eanos, FCEyN, Universidad de Buenos Aires and CONICET, Buenos Aires, Argentina
\item[$^{3}$] IFLP, Universidad Nacional de La Plata and CONICET, La Plata, Argentina
\item[$^{4}$] Instituto de Astronom\'\i{}a y F\'\i{}sica del Espacio (IAFE, CONICET-UBA), Buenos Aires, Argentina
\item[$^{5}$] Instituto de F\'\i{}sica de Rosario (IFIR) -- CONICET/U.N.R.\ and Facultad de Ciencias Bioqu\'\i{}micas y Farmac\'euticas U.N.R., Rosario, Argentina
\item[$^{6}$] Instituto de Tecnolog\'\i{}as en Detecci\'on y Astropart\'\i{}culas (CNEA, CONICET, UNSAM), and Universidad Tecnol\'ogica Nacional -- Facultad Regional Mendoza (CONICET/CNEA), Mendoza, Argentina
\item[$^{7}$] Instituto de Tecnolog\'\i{}as en Detecci\'on y Astropart\'\i{}culas (CNEA, CONICET, UNSAM), Buenos Aires, Argentina
\item[$^{8}$] International Center of Advanced Studies and Instituto de Ciencias F\'\i{}sicas, ECyT-UNSAM and CONICET, Campus Miguelete -- San Mart\'\i{}n, Buenos Aires, Argentina
\item[$^{9}$] Laboratorio Atm\'osfera -- Departamento de Investigaciones en L\'aseres y sus Aplicaciones -- UNIDEF (CITEDEF-CONICET), Argentina
\item[$^{10}$] Observatorio Pierre Auger, Malarg\"ue, Argentina
\item[$^{11}$] Observatorio Pierre Auger and Comisi\'on Nacional de Energ\'\i{}a At\'omica, Malarg\"ue, Argentina
\item[$^{12}$] Universidad Tecnol\'ogica Nacional -- Facultad Regional Buenos Aires, Buenos Aires, Argentina
\item[$^{13}$] University of Adelaide, Adelaide, S.A., Australia
\item[$^{14}$] Universit\'e Libre de Bruxelles (ULB), Brussels, Belgium
\item[$^{15}$] Vrije Universiteit Brussels, Brussels, Belgium
\item[$^{16}$] Centro Brasileiro de Pesquisas Fisicas, Rio de Janeiro, RJ, Brazil
\item[$^{17}$] Centro Federal de Educa\c{c}\~ao Tecnol\'ogica Celso Suckow da Fonseca, Petropolis, Brazil
\item[$^{18}$] Instituto Federal de Educa\c{c}\~ao, Ci\^encia e Tecnologia do Rio de Janeiro (IFRJ), Brazil
\item[$^{19}$] Universidade de S\~ao Paulo, Escola de Engenharia de Lorena, Lorena, SP, Brazil
\item[$^{20}$] Universidade de S\~ao Paulo, Instituto de F\'\i{}sica de S\~ao Carlos, S\~ao Carlos, SP, Brazil
\item[$^{21}$] Universidade de S\~ao Paulo, Instituto de F\'\i{}sica, S\~ao Paulo, SP, Brazil
\item[$^{22}$] Universidade Estadual de Campinas (UNICAMP), IFGW, Campinas, SP, Brazil
\item[$^{23}$] Universidade Estadual de Feira de Santana, Feira de Santana, Brazil
\item[$^{24}$] Universidade Federal de Campina Grande, Centro de Ciencias e Tecnologia, Campina Grande, Brazil
\item[$^{25}$] Universidade Federal do ABC, Santo Andr\'e, SP, Brazil
\item[$^{26}$] Universidade Federal do Paran\'a, Setor Palotina, Palotina, Brazil
\item[$^{27}$] Universidade Federal do Rio de Janeiro, Instituto de F\'\i{}sica, Rio de Janeiro, RJ, Brazil
\item[$^{28}$] Universidad de Medell\'\i{}n, Medell\'\i{}n, Colombia
\item[$^{29}$] Universidad Industrial de Santander, Bucaramanga, Colombia
\item[$^{30}$] Charles University, Faculty of Mathematics and Physics, Institute of Particle and Nuclear Physics, Prague, Czech Republic
\item[$^{31}$] Institute of Physics of the Czech Academy of Sciences, Prague, Czech Republic
\item[$^{32}$] Palacky University, Olomouc, Czech Republic
\item[$^{33}$] CNRS/IN2P3, IJCLab, Universit\'e Paris-Saclay, Orsay, France
\item[$^{34}$] Laboratoire de Physique Nucl\'eaire et de Hautes Energies (LPNHE), Sorbonne Universit\'e, Universit\'e de Paris, CNRS-IN2P3, Paris, France
\item[$^{35}$] Univ.\ Grenoble Alpes, CNRS, Grenoble Institute of Engineering Univ.\ Grenoble Alpes, LPSC-IN2P3, 38000 Grenoble, France
\item[$^{36}$] Universit\'e Paris-Saclay, CNRS/IN2P3, IJCLab, Orsay, France
\item[$^{37}$] Bergische Universit\"at Wuppertal, Department of Physics, Wuppertal, Germany
\item[$^{38}$] Karlsruhe Institute of Technology (KIT), Institute for Experimental Particle Physics, Karlsruhe, Germany
\item[$^{39}$] Karlsruhe Institute of Technology (KIT), Institut f\"ur Prozessdatenverarbeitung und Elektronik, Karlsruhe, Germany
\item[$^{40}$] Karlsruhe Institute of Technology (KIT), Institute for Astroparticle Physics, Karlsruhe, Germany
\item[$^{41}$] RWTH Aachen University, III.\ Physikalisches Institut A, Aachen, Germany
\item[$^{42}$] Universit\"at Hamburg, II.\ Institut f\"ur Theoretische Physik, Hamburg, Germany
\item[$^{43}$] Universit\"at Siegen, Department Physik -- Experimentelle Teilchenphysik, Siegen, Germany
\item[$^{44}$] Gran Sasso Science Institute, L'Aquila, Italy
\item[$^{45}$] INFN Laboratori Nazionali del Gran Sasso, Assergi (L'Aquila), Italy
\item[$^{46}$] INFN, Sezione di Catania, Catania, Italy
\item[$^{47}$] INFN, Sezione di Lecce, Lecce, Italy
\item[$^{48}$] INFN, Sezione di Milano, Milano, Italy
\item[$^{49}$] INFN, Sezione di Napoli, Napoli, Italy
\item[$^{50}$] INFN, Sezione di Roma ``Tor Vergata'', Roma, Italy
\item[$^{51}$] INFN, Sezione di Torino, Torino, Italy
\item[$^{52}$] Istituto di Astrofisica Spaziale e Fisica Cosmica di Palermo (INAF), Palermo, Italy
\item[$^{53}$] Osservatorio Astrofisico di Torino (INAF), Torino, Italy
\item[$^{54}$] Politecnico di Milano, Dipartimento di Scienze e Tecnologie Aerospaziali , Milano, Italy
\item[$^{55}$] Universit\`a del Salento, Dipartimento di Matematica e Fisica ``E.\ De Giorgi'', Lecce, Italy
\item[$^{56}$] Universit\`a dell'Aquila, Dipartimento di Scienze Fisiche e Chimiche, L'Aquila, Italy
\item[$^{57}$] Universit\`a di Catania, Dipartimento di Fisica e Astronomia ``Ettore Majorana``, Catania, Italy
\item[$^{58}$] Universit\`a di Milano, Dipartimento di Fisica, Milano, Italy
\item[$^{59}$] Universit\`a di Napoli ``Federico II'', Dipartimento di Fisica ``Ettore Pancini'', Napoli, Italy
\item[$^{60}$] Universit\`a di Palermo, Dipartimento di Fisica e Chimica ''E.\ Segr\`e'', Palermo, Italy
\item[$^{61}$] Universit\`a di Roma ``Tor Vergata'', Dipartimento di Fisica, Roma, Italy
\item[$^{62}$] Universit\`a Torino, Dipartimento di Fisica, Torino, Italy
\item[$^{63}$] Benem\'erita Universidad Aut\'onoma de Puebla, Puebla, M\'exico
\item[$^{64}$] Unidad Profesional Interdisciplinaria en Ingenier\'\i{}a y Tecnolog\'\i{}as Avanzadas del Instituto Polit\'ecnico Nacional (UPIITA-IPN), M\'exico, D.F., M\'exico
\item[$^{65}$] Universidad Aut\'onoma de Chiapas, Tuxtla Guti\'errez, Chiapas, M\'exico
\item[$^{66}$] Universidad Michoacana de San Nicol\'as de Hidalgo, Morelia, Michoac\'an, M\'exico
\item[$^{67}$] Universidad Nacional Aut\'onoma de M\'exico, M\'exico, D.F., M\'exico
\item[$^{68}$] Institute of Nuclear Physics PAN, Krakow, Poland
\item[$^{69}$] University of \L{}\'od\'z, Faculty of High-Energy Astrophysics,\L{}\'od\'z, Poland
\item[$^{70}$] Laborat\'orio de Instrumenta\c{c}\~ao e F\'\i{}sica Experimental de Part\'\i{}culas -- LIP and Instituto Superior T\'ecnico -- IST, Universidade de Lisboa -- UL, Lisboa, Portugal
\item[$^{71}$] ``Horia Hulubei'' National Institute for Physics and Nuclear Engineering, Bucharest-Magurele, Romania
\item[$^{72}$] Institute of Space Science, Bucharest-Magurele, Romania
\item[$^{73}$] Center for Astrophysics and Cosmology (CAC), University of Nova Gorica, Nova Gorica, Slovenia
\item[$^{74}$] Experimental Particle Physics Department, J.\ Stefan Institute, Ljubljana, Slovenia
\item[$^{75}$] Universidad de Granada and C.A.F.P.E., Granada, Spain
\item[$^{76}$] Instituto Galego de F\'\i{}sica de Altas Enerx\'\i{}as (IGFAE), Universidade de Santiago de Compostela, Santiago de Compostela, Spain
\item[$^{77}$] IMAPP, Radboud University Nijmegen, Nijmegen, The Netherlands
\item[$^{78}$] Nationaal Instituut voor Kernfysica en Hoge Energie Fysica (NIKHEF), Science Park, Amsterdam, The Netherlands
\item[$^{79}$] Stichting Astronomisch Onderzoek in Nederland (ASTRON), Dwingeloo, The Netherlands
\item[$^{80}$] Universiteit van Amsterdam, Faculty of Science, Amsterdam, The Netherlands
\item[$^{81}$] Case Western Reserve University, Cleveland, OH, USA
\item[$^{82}$] Colorado School of Mines, Golden, CO, USA
\item[$^{83}$] Department of Physics and Astronomy, Lehman College, City University of New York, Bronx, NY, USA
\item[$^{84}$] Michigan Technological University, Houghton, MI, USA
\item[$^{85}$] New York University, New York, NY, USA
\item[$^{86}$] University of Chicago, Enrico Fermi Institute, Chicago, IL, USA
\item[$^{87}$] University of Delaware, Department of Physics and Astronomy, Bartol Research Institute, Newark, DE, USA
\item[] -----
\item[$^{a}$] Max-Planck-Institut f\"ur Radioastronomie, Bonn, Germany
\item[$^{b}$] also at Kapteyn Institute, University of Groningen, Groningen, The Netherlands
\item[$^{c}$] School of Physics and Astronomy, University of Leeds, Leeds, United Kingdom
\item[$^{d}$] Fermi National Accelerator Laboratory, Fermilab, Batavia, IL, USA
\item[$^{e}$] Pennsylvania State University, University Park, PA, USA
\item[$^{f}$] Colorado State University, Fort Collins, CO, USA
\item[$^{g}$] Louisiana State University, Baton Rouge, LA, USA
\item[$^{h}$] now at Graduate School of Science, Osaka Metropolitan University, Osaka, Japan
\item[$^{i}$] Institut universitaire de France (IUF), France
\item[$^{j}$] now at Technische Universit\"at Dortmund and Ruhr-Universit\"at Bochum, Dortmund and Bochum, Germany
\end{description}

\section*{Acknowledgments}

\begin{sloppypar}
The successful installation, commissioning, and operation of the Pierre
Auger Observatory would not have been possible without the strong
commitment and effort from the technical and administrative staff in
Malarg\"ue. We are very grateful to the following agencies and
organizations for financial support:
\end{sloppypar}

\begin{sloppypar}
Argentina -- Comisi\'on Nacional de Energ\'\i{}a At\'omica; Agencia Nacional de
Promoci\'on Cient\'\i{}fica y Tecnol\'ogica (ANPCyT); Consejo Nacional de
Investigaciones Cient\'\i{}ficas y T\'ecnicas (CONICET); Gobierno de la
Provincia de Mendoza; Municipalidad de Malarg\"ue; NDM Holdings and Valle
Las Le\~nas; in gratitude for their continuing cooperation over land
access; Australia -- the Australian Research Council; Belgium -- Fonds
de la Recherche Scientifique (FNRS); Research Foundation Flanders (FWO),
Marie Curie Action of the European Union Grant No.~101107047; Brazil --
Conselho Nacional de Desenvolvimento Cient\'\i{}fico e Tecnol\'ogico (CNPq);
Financiadora de Estudos e Projetos (FINEP); Funda\c{c}\~ao de Amparo \`a
Pesquisa do Estado de Rio de Janeiro (FAPERJ); S\~ao Paulo Research
Foundation (FAPESP) Grants No.~2019/10151-2, No.~2010/07359-6 and
No.~1999/05404-3; Minist\'erio da Ci\^encia, Tecnologia, Inova\c{c}\~oes e
Comunica\c{c}\~oes (MCTIC); Czech Republic -- GACR 24-13049S, CAS LQ100102401,
MEYS LM2023032, CZ.02.1.01/0.0/0.0/16{\textunderscore}013/0001402,
CZ.02.1.01/0.0/0.0/18{\textunderscore}046/0016010 and
CZ.02.1.01/0.0/0.0/17{\textunderscore}049/0008422 and CZ.02.01.01/00/22{\textunderscore}008/0004632;
France -- Centre de Calcul IN2P3/CNRS; Centre National de la Recherche
Scientifique (CNRS); Conseil R\'egional Ile-de-France; D\'epartement
Physique Nucl\'eaire et Corpusculaire (PNC-IN2P3/CNRS); D\'epartement
Sciences de l'Univers (SDU-INSU/CNRS); Institut Lagrange de Paris (ILP)
Grant No.~LABEX ANR-10-LABX-63 within the Investissements d'Avenir
Programme Grant No.~ANR-11-IDEX-0004-02; Germany -- Bundesministerium
f\"ur Bildung und Forschung (BMBF); Deutsche Forschungsgemeinschaft (DFG);
Finanzministerium Baden-W\"urttemberg; Helmholtz Alliance for
Astroparticle Physics (HAP); Helmholtz-Gemeinschaft Deutscher
Forschungszentren (HGF); Ministerium f\"ur Kultur und Wissenschaft des
Landes Nordrhein-Westfalen; Ministerium f\"ur Wissenschaft, Forschung und
Kunst des Landes Baden-W\"urttemberg; Italy -- Istituto Nazionale di
Fisica Nucleare (INFN); Istituto Nazionale di Astrofisica (INAF);
Ministero dell'Universit\`a e della Ricerca (MUR); CETEMPS Center of
Excellence; Ministero degli Affari Esteri (MAE), ICSC Centro Nazionale
di Ricerca in High Performance Computing, Big Data and Quantum
Computing, funded by European Union NextGenerationEU, reference code
CN{\textunderscore}00000013; M\'exico -- Consejo Nacional de Ciencia y Tecnolog\'\i{}a
(CONACYT) No.~167733; Universidad Nacional Aut\'onoma de M\'exico (UNAM);
PAPIIT DGAPA-UNAM; The Netherlands -- Ministry of Education, Culture and
Science; Netherlands Organisation for Scientific Research (NWO); Dutch
national e-infrastructure with the support of SURF Cooperative; Poland
-- Ministry of Education and Science, grants No.~DIR/WK/2018/11 and
2022/WK/12; National Science Centre, grants No.~2016/22/M/ST9/00198,
2016/23/B/ST9/01635, 2020/39/B/ST9/01398, and 2022/45/B/ST9/02163;
Portugal -- Portuguese national funds and FEDER funds within Programa
Operacional Factores de Competitividade through Funda\c{c}\~ao para a Ci\^encia
e a Tecnologia (COMPETE); Romania -- Ministry of Research, Innovation
and Digitization, CNCS-UEFISCDI, contract no.~30N/2023 under Romanian
National Core Program LAPLAS VII, grant no.~PN 23 21 01 02 and project
number PN-III-P1-1.1-TE-2021-0924/TE57/2022, within PNCDI III; Slovenia
-- Slovenian Research Agency, grants P1-0031, P1-0385, I0-0033, N1-0111;
Spain -- Ministerio de Ciencia e Innovaci\'on/Agencia Estatal de
Investigaci\'on (PID2019-105544GB-I00, PID2022-140510NB-I00 and
RYC2019-027017-I), Xunta de Galicia (CIGUS Network of Research Centers,
Consolidaci\'on 2021 GRC GI-2033, ED431C-2021/22 and ED431F-2022/15),
Junta de Andaluc\'\i{}a (SOMM17/6104/UGR and P18-FR-4314), and the European
Union (Marie Sklodowska-Curie 101065027 and ERDF); USA -- Department of
Energy, Contracts No.~DE-AC02-07CH11359, No.~DE-FR02-04ER41300,
No.~DE-FG02-99ER41107 and No.~DE-SC0011689; National Science Foundation,
Grant No.~0450696, and NSF-2013199; The Grainger Foundation; Marie
Curie-IRSES/EPLANET; European Particle Physics Latin American Network;
and UNESCO.
\end{sloppypar}

}


\begin{thebibliography}{99}

\small
\setlength{\itemsep}{1pt}

\bibitem{PierreAuger:2017pzq}
A.~Aab \textit{et al.} [Pierre Auger],
Science \textbf{357}, no.6537, 1266-1270 (2017)

\bibitem{PierreAuger:2024fgl}
A.~A.~Halim \textit{et al.} [Pierre Auger],
Astrophys. J. \textbf{976}, no.1, 48 (2024)

\bibitem{PierreAuger:2014sui}
A.~Aab \textit{et al.} [Pierre Auger],
Phys. Rev. D \textbf{90}, no.12, 122005 (2014)

\bibitem{PierreAuger:2014gko}
A.~Aab \textit{et al.} [Pierre Auger],
Phys. Rev. D \textbf{90}, no.12, 122006 (2014)

\bibitem{PierreAuger:2024nzw}
A.~Abdul Halim \textit{et al.} [Pierre Auger],
Phys. Rev. D \textbf{111}, no.2, 022003 (2025)

\bibitem{PierreAuger:2024flk}
A.~Abdul Halim \textit{et al.} [Pierre Auger],
Phys. Rev. Lett. \textbf{134}, no.2, 021001 (2025)

\bibitem{Smida:2017zmh}
R.~\v{S}m\'\i{}da [Pierre Auger],
PoS \textbf{ICRC2017}, 390 (2018)

\bibitem{Pkala:2020cro}
J.~Pekala [Pierre Auger],
PoS \textbf{ICRC2019}, 380 (2020)

\bibitem{ICRC25:Matteo}
M.~Conte \textit{et al.} [Pierre Auger], these proceedings.

\bibitem{ICRC25:Belen}
B.~Andrada \textit{et al.} [Pierre Auger], these proceedings.

\bibitem{ICRC25:Bjarni}
B.~Pont \textit{et al.} [Pierre Auger], these proceedings.

\bibitem{PierreAuger:2023clx}
A.~Abdul Halim \textit{et al.} [Pierre Auger],
PoS \textbf{ICRC2023}, 343 (2023)

\bibitem{ICRC25:Martina}
M.~Bohacova \textit{et al.} [Pierre Auger], these proceedings.

\bibitem{PierreAuger:2023yab}
A.~Abdul Halim \textit{et al.} [Pierre Auger],
JINST \textbf{18}, no.10, P10016 (2023)

\bibitem{PierreAuger:2020hrz}
A.~Aab \textit{et al.} [Pierre Auger],
JINST \textbf{16}, no.04, P04003 (2021)

\bibitem{ICRC25:Marina}
M.~Scornavacche \textit{et al.} [Pierre Auger], these proceedings.

\bibitem{ICRC25:Joaquin}
J.~de Jesus \textit{et al.} [Pierre Auger], these proceedings.

\bibitem{ICRC25:Paul}
P.~Filip \textit{et al.} [Pierre Auger], these proceedings.

\bibitem{ICRC25:Simon}
S.~Strähnz \textit{et al.} [Pierre Auger], these proceedings.

\bibitem{ICRC25:Darko}
M.~Stadelmaier \textit{et al.} [Pierre Auger], these proceedings.

\bibitem{ICRC25:Harm}
H.~Schoorlemmer \textit{et al.} [Pierre Auger], these proceedings.

\bibitem{ICRC25:Tobias}
T.~Schulz \textit{et al.} [Pierre Auger], these proceedings.

\end{thebibliography}
\end{document}